# Brownian markets

Roumen Tsekov
Department of Physical Chemistry, University of Sofia, 1164 Sofia, Bulgaria

Financial market dynamics is rigorously studied via the exact generalized Langevin equation. Assuming market Brownian self-similarity, the market return rate memory and autocorrelation functions are derived, which exhibit an oscillatory-decaying behavior with a long-time tail, similar to empirical observations. Individual stocks are also described via the generalized Langevin equation. They are classified by their relation to the market memory as heavy, neutral and light stocks, possessing different kinds of autocorrelation functions.

In 1900 Bachelier, a student of Poincaré at that time, has published his doctoral thesis entitled "Théorie de la speculation" [1], where he has developed the mathematical theory of the Brownian motion five years before the famous Einstein's paper [2] has come out to explain its physics. Bachelier introduced also the geometric Brownian motion (GBM), which is the background of the modern Black-Scholes option pricing model [3] and the most powerful tool for quantitative description of stock market fluctuations [4-6]. Thus, he initiated the econophysics, aiming to explain the complex economical phenomena by physical laws [7-9].

According to GBM, the market fluctuations obey a stochastic differential equation

$$dM = \mu M dt + \sigma M dW \tag{1}$$

where $M$ is the market prize, $\mu$ is the market drift, $t$ is time, $\sigma$ is the market volatility and $W$ is a random Wiener process. As is seen, the noise in Eq. (1) is multiplicative. In finances, the stochastic product $MdW$ is traditionally treated via the Ito lemma [10] but there are also other definitions proposed in the literature for handling of this peculiarity [11, 12]. Equation (1) describes GBM without memory, while the financial markets are driven by people, who possess ability to remember. Hence, the GBM model (1) is oversimplified and requires a generalization, which is the scope of the present paper. An explicit expression for the market return rate memory function is derived based on the Brownian self-similarity [13]. This concept is already applied to hydrodynamic memory [14] and it corresponds to the simplest Hermitian dynamics, governed by an infinite-dimensional hyperspherical Hilbert space [15]. New models of more complex Brownian self-similarity are also proposed here, which are based on scaling and fractals.

In the frames of the classical mechanics the evolution of an observable $R(t)$, being a function of momentums and coordinates of all the particles in the Universe, is governed by the following dynamic equation

$$dR(t) = i\hat{L}R(t)dt \tag{2}$$

where $i\hat{L} = \{\cdot, H\}$ is the global Liouville operator with $H$ being the Universe Hamiltonian. The latter takes into account all the interactions in the whole Universe, including the human activities as well. Equation (2) is an alternative presentation of the Newtonian laws from the classical mechanics. The formal solution of Eq. (2) can be written in the form

$$R(t) = \exp(i\hat{L}t)R \tag{3}$$

where $R \equiv R(0)$ is the initial value of the observable. This exact solution is, however, useless since no one is able to define precisely the Universe Liouvillian and even its approximations will not make the problem easier since Eq. (3) involves infinite number of differentiations. In addition, the dependence of $R$ on particles coordinates and momentums is usually unknown.

Obviously, we are not able to describe rigorously the evolution of the whole Universe but our interest is concentrated solely on the description of a very small part of it, particularly, the prize $M$ of a market. Of course, the latter is influenced by many processes in the Nature but some of them are important, while others are meaningless. Hence, the basic idea in statistical physics is to introduce a projection operator $\hat{P}$, which focuses the observation on the variable $R$. Evidently, the projector satisfies idempotence $\hat{P}^2 = \hat{P}$ and a possible definition of the projection operator reads

$$\hat{P}X \equiv R<RX>/<R^2> \tag{4}$$

where $<\cdot>$ denotes a statistical average. As is seen, the operator $\hat{P}$ from Eq. (4) projects the effect of $X$ on $R$ via the correlation $<RX>$ between these two quantities. If they are statistically independent and zero centered than $<RX> = <R><X> = 0$, and the evolution of $R$ will not be affected by $X$ in an average sense. On the other hand the projector (4) preserves completely the information about $R$ since $\hat{P}R = R$.

In the physical literature a general integral representation for the exponential operator form Eq. (3) is proposed, which is the base of the Mori-Zwanzig formalism [16, 17],

$$\exp(i\hat{L}t) \equiv \int_0^t \exp(i\hat{L}s)]\hat{P}i\hat{L}\exp[(1-\hat{P})i\hat{L}(t-s)]ds + \exp[(1-\hat{P})i\hat{L}t] \tag{5}$$

Applying this integral identity on the initial velocity $i\hat{L}R$ and using Eq. (4) leads to the following dynamic equation equivalent to Eq. (2)

$$\frac{dR(t)}{dt} = -\int_0^t \frac{<F(t)F(s)>}{<R^2>} R(s)ds + F(t) \tag{6}$$

where the random fluctuation force is introduced via $F(t) \equiv \exp[(1-\hat{P})i\hat{L}t]i\hat{L}R$. The benefit of the exact Mori-Zwanzig presentation (5) and the corresponding generalized Langevin equation GLE (6) is the separation of the entire interaction into two universal forces, dissipation and fluctuation ones, governing the evolution on a macroscopic level. The integral in Eq. (6) represents the dissipation force. The fluctuation-dissipation theorem is also emphasized in Eq. (6) by the fact that the memory kernel in this integral is proportional to the autocorrelation function of the fluctuation force. In addition, the rigorous definition of $F(t)$ above proves the exact relations $<F(t)>=0$ and $<F(t)R>=0$, where the latter means that there is no correlation between the Langevin force at a given moment and the observable at the beginning. Using these relations one can derive, via multiplying Eq. (6) by $R$ and taking an average value, an integro-differential equation

$$\frac{dC_{RR}(\tau)}{d\tau} = -\int_0^\tau \frac{C_{FF}(\tau-s)}{C_{RR}(0)} C_{RR}(s)ds \tag{7}$$

for the observable autocorrelation function $C_{RR}(\tau) \equiv <R(\tau)R>$ as related to the Langevin force autocorrelation function $C_{FF}(\tau) \equiv <F(\tau)F>$.

Applying a standard Laplace transformation to Eq. (7) results in the following image expression (the Laplace images are denoted by tilde)

$$\tilde{C}_{RR}(p) = C_{RR}(0)/[p + \tilde{C}_{FF}(p)/C_{RR}(0)] \tag{8}$$

where $p$ is the Laplace transformation variable. As is seen the autocorrelation function of the Langevin force $C_{FF}$ determines uniquely the autocorrelation function of the observable. The derivation of the equations above is general and can be applied to arbitrary observable, which is stationary and zero centered. A very popular model for the fluctuation Langevin force is the white noise with a constant spectral density, $\tilde{C}_{FF}(p) = <R^2>/\tau_R$, where $\tau_R$ is the correlation time of the observable $R(t)$. In this case the inverse image of Eq. (8) represents an exponentially decaying autocorrelation function

$$C_{RR}(\tau) = <R^2> \exp(-\tau/\tau_R) \tag{9}$$

which is typical for stationary Gaussian Markov processes according to the Doob theorem [18]. The stochastic differential equation corresponding to the white noise Langevin force reads

$$dR(t) = -R(t)dt/\tau_R + \sqrt{2<R^2>/\tau_R}dW \qquad (10)$$

The application of GLE to stock markets [19] requires a proper definition of the observable $R(t)$. Since we are looking for a zero centered ($<R>=0$) stationary variable with a constant dispersion $<R^2>$, following Bachelier a proper candidate is the market return rate fluctuation

$$R(t) = d\ln M/dt - \mu \qquad (11)$$

where $M$ is the market prize and the drift $\mu$ is its mean rate of return. In physics, stationary processes are usually the rates of change of some quantities, e.g. velocity of a molecule, etc. For this reason the variable in Eq. (11) is not simply proportional to the market prize $M$ but to its relative rate of change. If the time $t$ is much larger than the relaxation time $\tau_R$ one can neglect the left-hand-side of Eq. (10) and thus it simplifies to $R(t)dt = \sqrt{2<R^2>\tau_R}dW$. Introducing here Eq. (11) results in a stochastic differential equation for the market prize

$$dM = \mu Mdt + \sqrt{2<R^2>\tau_R}MdW \qquad (12)$$

Comparing now this equation with Eq. (1) unveils an expression relating the correlation time $\tau_R$ by the market volatility $\sigma$ ($\sigma^2/2$ is the return diffusion constant) and dispersion $<R^2>$ of the market return fluctuations, being proportional to the market temperature,

$$\tau_R = \sigma^2/2<R^2> \qquad (13)$$

If the dimensionless return rate $d\ln M/d\mu t$ obeys the Poisson law than $<R^2>=\mu^2$ and the correlation time from Eq. (13) acquires the simple form $\tau_R = \sigma^2/2\mu^2$. However, since according to Eq. (12) the return rate fluctuation $R(t)$ is proportional to a white noise, which is only Gaussian [20], it is reasonable to accept that the return rate fluctuations are Gaussian as well. In reality, the market return rate is fat-tail distributed [21-27], which indicates that the white noise approximation is rough and the memory effects due to speculation/regulation are very important.

The analysis above shows that Eq. (1) is valid only for the case of lack of memory and at large time $t > \tau_R$. A general way to determine the return rate autocorrelation and memory functions is to assume Brownian self-similarity of the market. According to this model [13] the autocorrelation functions of the observable and its conjugated Langevin force are the same

$$C_{FF}(\tau)/C_{FF}(0) = C_{RR}(\tau)/C_{RR}(0) \qquad (14)$$

which can be translated in common words as "a market is driven by the market itself". Strictly speaking the assumption (14) implies a Hermitian dynamics in an infinite-dimensional hyperspherical Hilbert space [15]. Such proportionality between forces and flows can be also seen in the linear non-equilibrium thermodynamics. Combining Eq. (14) with Eq. (8) results in the following expression for the Laplace image of the return rate autocorrelation function

$$\tilde{C}_{RR}(p) = <R^2> \tau_R [\sqrt{1+(\tau_R p/2)^2} - \tau_R p/2] \qquad (15)$$

where the correlation time equals to $\tau_R \equiv \sqrt{<R^2>/<F^2>}$. The inverse Laplace transformation of Eq. (15) leads straightforward to return rate autocorrelation function [13]

$$C_{RR}(\tau)/<R^2> = (\tau_R/\tau)J_1(2\tau/\tau_R) = \Lambda_1(2\tau/\tau_R) \qquad (16)$$

where $J_1$ is the Bessel function of first kind and first order and $\Lambda_1$ is a lambda function. The latter is universal oscillatory-decaying one, whose amplitude exhibits asymptotically a long-time tail falling as $(\tau_R/\tau)^{3/2}$. Its plot in Fig. 1 shows the existence of a sequence of correlations and anti-correlations of the market return rate being particularly responsible for the Elliott waves.

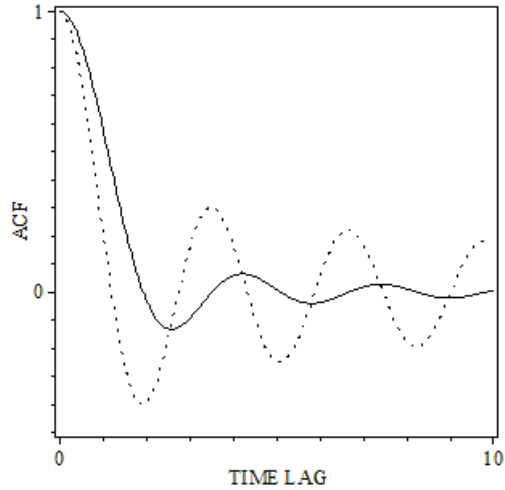

**Fig. 1** The dependence of the return rate autocorrelation functions $\Lambda_1$ from Eq. (16) (solid line) and $\Lambda_0$ from Eq. (21) (dotted line) as a function of the dimensionless time lag $\tau/\tau_R$.

The autocorrelation function (16) is known in physics as the Rubin model [28]. Similar autocorrelation functions are empirically detected in the Dow Jones Industrial Average index [29] and DAX [30, 31], for instance. According to Eqs. (14) and (16) the corresponding memory kernel is $C_{FF}/<R^2> = \Lambda_1(2\tau/\tau_R)/\tau_R^2$ and thus Eq. (6) becomes fully specified. Since the Langevin force is not a white noise anymore, the return rate fluctuations $R(t)$ are not restricted to be Gaussian, in contrast to Eq. (1). Hence, Eq. (6) describes completely the market stochastic dynamics by accounting for the red color of the market noise $F(t)$ and the resultant memory effect. Further, GLE (6) can be employed for derivation of Fokker-Planck equations [32], describing the probability density evolution in another desired space related to the market return rate.

GLE can be applied also for description of the return rate $r(t) = d\ln S/dt - \mu_S$ of an individual stock, where $S$ and $\mu_S$ are the stock prize and drift, respectively,

$$\frac{dr(t)}{dt} + \int_0^t \frac{C_{ff}(t-s)}{<r^2>} r(s)ds = f(t) \tag{17}$$

The return rate autocorrelation function can be expressed analogically to Eq. (8) in the form

$$\tilde{C}_{rr}(p)/\tilde{C}_{rr}(0) = 1/[\tau_r p + \tilde{C}_{ff}(p)/\tilde{C}_{ff}(0)] \tag{18}$$

The stock correlation time is introduced here via the standard definition $\tau_r \equiv \tilde{C}_{rr}(0)/<r^2>$ and accounts for inertial effects as well. Since the observed stock interacts with other stocks from the market the concept of Brownian self-similarity cannot be applied in its simplest form (14). In a first approximation one could accept, however, that the stochastic forces acting on individual stocks have the same memory as the common market noise $F(t)$ and the market itself. Hence, the spectral density $\tilde{C}_{ff}$ of the Langevin force autocorrelation function can be expressed in the already derived form

$$\tilde{C}_{ff}(p)/\tilde{C}_{ff}(0) = \sqrt{1+(\tau_R p/2)^2} - \tau_R p/2 \tag{19}$$

where $\tau_R$ is the correlation time of the market return rate fluctuations. As is seen, the introduction of Eq. (19) is somehow heuristic but we hope that some speculations are allowed anyway in a theory of speculations. Substituting now Eq. (19) in Eq. (18) yields

$$\tilde{C}_{rr}(p) = <r^2>\tau_r/[\tau_r p + \sqrt{1+(\tau_R p/2)^2} - \tau_R p/2] \tag{20}$$

According to Eq. (20) the stock return rate autocorrelation function involves two relaxation times, $\tau_r$ and $\tau_R$. Since $\tau_r$ is the stock correlation time, which scales the Laplace transform variable $p$, it is reasonable to introduce a dimensionless parameter $\theta \equiv \tau_R/\tau_r$, which can help to classify stocks. One can call a stock heavy if $0 \leq \theta < 2/3$ since its correlation time is larger than the market return relaxation time. Hence, such stocks are inert and not affected substantially by the faster market fluctuations. Indeed, according to Eq. (20) the heaviest stock with $\theta = 0$ possesses an exponential autocorrelation function $C_{rr}/<r^2> = \exp(-\tau/\tau_r)$, which corresponds to the case of lack of collective memory. Obviously, heavy stocks are appropriate for investment strategies due to their long correlation time. Stocks in the range $2/3 \leq \theta < 4/3$ can be called neutral, since they follow more or less the market fluctuations. Indeed, the most neutral stock with $\theta = 1$ possesses from Eq. (20) a return rate autocorrelation function identical to that of the market return rate autocorrelation from Eq. (16), $C_{rr}/<r^2> = \Lambda_1(2\tau/\tau_r)$. The light stocks correspond to $4/3 \leq \theta < 2$ and their correlation time is shorter than $\tau_R$, while ultra light stocks with $\theta \geq 2$ possess a maximum of $\tilde{C}_{rr}$. According to Eq. (20) an example for autocorrelation function of a light stock with $\theta = 2$ is the Bessel function of first kind and zero order [13]

$$C_{rr}/<r^2> = J_0(\tau/\tau_r) = \Lambda_0(2\tau/\tau_R) \tag{21}$$

The plot of this autocorrelation function in Fig. 1 shows a strong periodicity, which indicates essential memory effects. Additionally, $\Lambda_0$ from Eq. (21) exhibits a longer time tail $(\tau_R/\tau)^{1/2}$ than $\Lambda_1$. Light stocks are suitable for traders, since they exhibit cyclic behavior due to strong collective memory effects from the market environment, but these correlations are short-living [7].

As was shown, the Brownian self-similarity concept requires existence of an additional relationship between the spectral densities $\tilde{C}_{ff}(p)/\tilde{C}_{ff}(0)$ and $\tilde{C}_{rr}(p)/\tilde{C}_{rr}(0)$ of the autocorrelation functions of the Langevin force and its conjugated observable. Introducing such a relationship in the exact Eq. (18) determines uniquely the observable and Langevin force autocorrelation functions as well as the memory kernel in Eq. (17). Hence, the Brownian self-similarity is a powerful tool for stochastic modeling of many diffusive processes in the Nature. As was mentioned before, Eq. (19) is a first order approximation, since it does not account for the feedback of the observed stock to the local environment. A way to improve the present theory is to employ Brownian self-similarity again but in a more general form. Thus, one can express the Langevin force spectral density, for instance, via scaling or fractional laws

$$\tilde{C}_{ff}(p)/\tilde{C}_{ff}(0) = \tilde{C}_{rr}(\theta p)/\tilde{C}_{rr}(0) \qquad \tilde{C}_{ff}(p)/\tilde{C}_{ff}(0) = [\tilde{C}_{rr}(p)/\tilde{C}_{rr}(0)]^\theta \tag{22}$$

which are inspired from the previous analysis. Both expressions lead to a white noise for the heaviest stock with $\theta = 0$ and to an analog of Eq. (14) for the most neutral stock with $\theta = 1$. Equations (22) describe more complex examples of Brownian self-similarity and can be employed in Eq. (18) for more sophisticated description of the stochastic dynamics of stocks.

The present paper describes financial markets from the viewpoint of the classical mechanics but GLE is even more general [33]. It is known that the best description of the Universe is given by the quantum mechanics and, if the analysis can be performed starting from the latter, one would arrive at quantum Brownian markets [34]. An interesting detail in this respect is achieved by employing of a differential Brownian self-similarity model $d\tilde{C}_{FF}/dp = \tilde{C}_{RR}(p)/\tau_R$. Introducing it in Eq. (8) and solving the resulting differential equation yield

$$\tilde{C}_{RR}(0)/\tilde{C}_{RR}(p) = -1 - W_{-1}[-2\exp(-2-\tau_R p)] = \tilde{C}_{FF}(p)/\tilde{C}_{FF}(0) + \tau_R p \qquad (23)$$

where $W_{-1}$ is a Lambert W-function. The market noise $F(t)$, corresponding to Eq. (23), is blue. What is exciting in Eq. (23) is that a very similar functional dependence is detected recently in the description of a Gaussian wave packet spreading in quantum Brownian motion [35]. Perhaps, the quantum market effects could also be accounted for in this way.